\documentclass[12pt]{article}

\begin{document}\begin{flushright}\thispagestyle{empty}

OUT--4102--92\\
IPPP/04/11\\
DCPT/04/22\\
\end{flushright}
\begin{center}{
                                                    \Large\bf
Comparison of the Gottfried and Adler sum rules\\
within the large-$N_c$ expansion
                                                    }\vglue 10mm{\large{\bf
D.~J.~Broadhurst$^{a,1)}$,
A.~L.~Kataev$^{b,2)}$
and
C.~J.~Maxwell$^{c,3)}$                              }}

$^{a)}$ Department of Physics and Astronomy, Open University,
Milton Keynes MK7 6AA, UK

$^{b)}$
Institute for Particle Physics Phenomenology, University
of Durham, DH1 3LE, UK and \\
Institute for Nuclear Research of the Academy of Sciences of Russia,
117312, Moscow, Russia

$^{c)}$ Institute for Particle Physics Phenomenology, University
of Durham, DH1 3LE, UK

\end{center}
{\bf Abstract}\quad
The Adler sum rule for deep inelastic neutrino scattering
measures the isospin of the nucleon and is hence exact.
By contrast, the corresponding Gottfried sum rule for charged lepton
scattering was based merely on a valence picture and is modified
both by perturbative and by non-perturbative effects.
Noting that the known perturbative corrections to two-loop order
are suppressed by a factor $1/N_c^2$, relative
to those for higher moments, we propose that this suppression
persists at higher orders and also
applies to higher-twist effects. Moreover,
we propose that the {\em differences} between the corresponding
radiative corrections to higher non-singlet moments in
charged-lepton and neutrino deep inelastic scattering are suppressed
by $1/N_c^2$, in all orders of perturbation theory.
For the first moment, in the Gottfried sum rule,
the substantial discrepancy between the measured value and the valence-model
expectation may be attributed to an intrinsic isospin
asymmetry in the nucleon sea, as is indeed the case
in a chiral-soliton model, where the discrepancy {\em persists}
in the limit $N_c\rightarrow\infty$.



\vfill\footnoterule\noindent
$^1$) D.Broadhurst@open.ac.uk\\
$^2$) Kataev@ms2.inr.ac.ru\\
$^3$) C.J.Maxwell@durham.ac.uk

\newpage\setcounter{page}{1}

\section{Introduction}

Alone among the various sum rules of deep inelastic scattering (DIS) the isospin
Adler sum rule~\cite{Adler} has the special feature that its
quark-parton model expression
\begin{eqnarray}
I_A&\equiv&\int_0^1\frac{{\rm d}x}{x}
\left[F_2^{\nu p}(x,Q^2)-F_2^{\nu n}(x,Q^2)\right]\\
\nonumber
&=&2\int_0^1{\rm d}x\left(u(x)-d(x)-\overline{u}(x)+\overline{d}(x)\right)
=4I_3=2
\end{eqnarray}
coincides with its QCD extension and receives neither perturbative
nor non-perturbative corrections
(for a discussion, see Ref.\cite{Dokshitzer:1995qm}).
Moreover, this sum rule is supported by the existing neutrino--nucleon
DIS data, which show no significant $Q^2$ variation in the range
$2~{\rm GeV}^2\leq Q^2\leq 30~{\rm GeV}^2$ and give~\cite{Allasia:1985hw}
\begin{equation}
I_A^{\rm exp}=2.02\pm 0.40\;.
\end{equation}
Though the error-bars are quite large, the precision could
in principle be improved by future $\nu N$ DIS experiments
at neutrino factories (for discussion
of such a program, see Ref.\cite{Mangano:2001mj}).

Within the quark-parton model, the corresponding isospin sum rule
in the case of charged-lepton--nucleon DIS has the form
\begin{eqnarray}
\label{GQPM}
I_G(Q^2)&\equiv&\int_0^1\frac{{\rm d}x}{x}\left[
F_2^{\ell p}(x,Q^2)-F_2^{\ell n}(x,Q^2)\right]
\\ \nonumber
&=&\frac{1}{3}\int_0^1{\rm d}x
\left(u(x)-d(x)+\overline{u}(x)-\overline{d}(x)\right)
\\ \nonumber
&=&\frac{1}{3}-\frac{2}{3}\int_0^1{\rm d}x
\left(\overline{d}(x)-\overline{u}(x)\right)
\;.
\end{eqnarray}
If the nucleon sea were flavour symmetric, with
$\overline{u}(x)=\overline{d}(x)$,
we would obtain the original Gottfried sum rule~\cite{Gottfried:1967kk},
$I_G=\frac13$, in strong disagreement with the most detailed
analysis of muon--nucleon DIS data, by the NMC collaboration, which gave
the following result~\cite{Arneodo:1994sh}:
\begin{equation}
\label{NMC}
I_G(Q^2=4~\rm{GeV}^2)=0.235\pm 0.026\;.
\end{equation}

In contrast to the Adler sum rule, the original quark-parton model expression
for the Gottfried sum rule is modified by perturbative QCD contributions,
analyzed numerically at the $\alpha_s^2$-level in Ref.\cite{Kataev:2003en}.
These corrections turn out to be small and cannot be responsible for the
significant discrepancy between $I_G$ and the naive expectation of $\frac13$.
This discrepancy can be associated with the existence of non-perturbative
effects in the nucleon sea, which generate light-quark flavour asymmetry,
and lead to the inequality $\overline{u}(x,Q^2)<\overline{d}(x,Q^2)$
over significant ranges of the Bjorken variable $x$ (for reviews, see
Refs.\cite{Kumano:1997cy,Garvey:2001yq,Kataev:2003xp}).

In this paper we examine the QCD corrections to the moments
of parton-model densities, for non-singlet neutrino and charged-lepton DIS,
with the $N=1$ moments corresponding to the Adler and Gottfried sum rules,
and comment upon a striking feature which they exhibit in the large-$N_c$
limit~\cite{'tHooft:1973jz} at the two-loop level.

\section{Radiative corrections at large $N_c$}

First we present an analytical result for the two-loop
radiative correction that was evaluated numerically
in Ref.\cite{Kataev:2003en}
and then comment on its structure as $N_c\rightarrow\infty$.

\subsection{Analytical two-loop correction to the Gottfried sum rule}

Following Ref.\cite{Kataev:2003en}, we write the radiative corrections
to the $N=1$ non-singlet charged-lepton moment of Eq.~(\ref{GQPM}),
in the case of light-quark flavour symmetry, as
\begin{equation}
\label{RG}
I_G=A(\alpha_s){C}^{(\ell)}(\alpha_s)\;,
\end{equation}
with an anomalous-dimension term
\begin{eqnarray}
\label{AD}
A(\alpha_s)&=&1+\frac{1}{8}\frac{\gamma_1^{N=1}}{\beta_0}
\left(\frac{\alpha_s}{\pi}\right) \\ \nonumber
&+&\frac{1}{64}\left(\frac{1}{2}\frac{(\gamma_1^{N=1})^2}{\beta_0^2}
-\frac{\gamma_1^{N=1}\beta_1}{\beta_0^2}+\frac{\gamma_2^{N=1}}{\beta_0}\right)
\left(\frac{\alpha_s}{\pi}\right)^2+O(\alpha_s^3)\;,
\end{eqnarray}
where $\beta_0$ and $\beta_1$ are the first two scheme-independent
coefficients of the QCD $\beta$-function, namely
\begin{eqnarray}
\beta_0&=&\left(\frac{11}{3}C_A-\frac{2}{3}N_F\right) \\
\beta_1&=&\left(\frac{34}{3}C_A^2-2C_FN_F-\frac{10}{3}C_AN_F\right)\;,
\end{eqnarray}
with $N_F$ active flavours and
Casimir operators $C_F=(N_c^2-1)/(2N_c)$ and $C_A=N_c$,
in the fundamental and adjoint representations of SU$(N_c)$.

The one-loop anomalous dimension vanishes and the leading correction
from the two-loop result of Ref.\cite{Ross:1978xk}, confirmed
in Ref.\cite{Curci:1980uw}, has the form
\begin{equation}
\label{gamma1}
\gamma_1^{N=1}=-4(C_F^2-C_FC_A/2)\left[13+8\zeta(3)-12\zeta(2)\right]
\approx2.557552\;,
\end{equation}
with two conspicuous features:
\begin{itemize}
\item
the appearance of
$\zeta(2)=\pi^2/6$, which is absent from even non-singlet moments of the
charged-lepton--nucleon structure function $F_2$, and from odd moments
of the corresponding neutrino--nucleon structure function, but occurs
at odd moments for charged-lepton scattering, and at even moments
for neutrino scattering, by analytic continuation in $N$ of results from
QCD Feynman diagrams~\cite{Curci:1980uw};
\item the distinctive non-planar colour-factor, $(C_F^2-C_FC_A/2)=O(N_c^0)$,
which exhibits an $O(1/N_c^2)$ suppression at large-$N_c$, in comparison
with the individual weights $C_F^2$ and $C_FC_A$, which are associated with
planar two-loop diagrams that do not show this large-$N_c$ cancellation at
two loops~\cite{Curci:1980uw} for moments $N>1$. Nor is there any sign
of such large-$N_c$ cancellation in the three-loop results
of~\cite{Larin:1993vu}, obtained for even moments.
\end{itemize}

The second factor in Eq.~(\ref{RG}) comes from radiative corrections
to the coefficient function, of the form
\begin{equation}
\label{coeff}
C^{(\ell)}(\alpha_s)=
\frac{1}{3}\left[1+C_1^{(\ell)N=1}\left(\frac{\alpha_s}{\pi}\right)
+ C_2^{(\ell)N=1}\left(\frac{\alpha_s}{\pi}\right)^2+O(\alpha_s^3)\right]
\end{equation}
with a vanishing one-loop term, $C_1^{(\ell)N=1}=0$~\cite{Bardeen:1978yd}.
The scheme-independent two-loop
coefficient $C_2^{(\ell)N=1}$
can be defined through the general non-singlet Mellin moment
of charged-lepton--nucleon ($\ell$) DIS scattering
\begin{equation}
\label{moment}
C_2^{(\ell)N}=3\int_0^1{\rm d}x\left[C^{(2),(+)}(x,1)+C^{(2),(-)}(x,1)\right]
x^{N-1}
\end{equation}
taken at $N=1$, where the expressions for the functions $C^{(2),(-)}(x,1)$
and $C^{(2),(+)}(x,1)$ were given in Ref.\cite{vanNeerven:1991nn}
and confirmed later with the help of another technique in
Ref.\cite{Moch:1999eb}. The ``$1$'' in the argument of these functions denotes
the choice of renormalization scale ${\mu}^{2}={Q}^{2}$, where
$\mu^2$ is associated to the
$\overline{{\rm MS}}$-scheme and the coupling
$\alpha_s$ is evaluated at $Q^2$.

Explicit numerical integration of the $N=1$ moment of Eq.~(\ref{moment})
gave the result~\cite{Kataev:2003en}
\begin{equation}
\label{result}
C_2^{(\ell)N=1}= 3.695C_F^2-1.847C_FC_A\;,
\end{equation}
with a contribution from the colour factor $C_FN_F$ which was consistent
with zero, to the accuracy of that numerical work. At the time,
the approximate emergence in Eq.~(\ref{result})
of the same non-planar structure
$(C_F^2-C_FC_A/2)$, already observed in the two-loop $N=1$
anomalous dimension coefficient of Eq.~(\ref{gamma1}), went unremarked.
Now we are able to derive an exact result, by comparing the charged-lepton
moments~(\ref{moment}) with the corresponding non-singlet moments
of the $F_2$ structure function for neutrino--nucleon ($\nu$) DIS,
which can also be expressed in terms of the functions
${C}^{(2),(-)}(x,1)$ and $C^{(2),(+)}(x,1)$,
but now in the combination
\begin{equation}
\label{nu}
{C}_{2}^{(\nu)N}=\frac{1}{2}\int_0^1{\rm d}x\left[{C}^{(2),(+)}(x,1)
-{C}^{(2),(-)}(x,1)\right]{x}^{N-1}\;.
\end{equation}

To obtain an analytic expression for the
correction~(\ref{moment}) to the Gottfried
sum rule we remark that the $N=1$ case of the moment~(\ref{nu})
corresponds to the Adler sum rule, which is free of radiative corrections.
Hence, ${C}_{2}^{(\nu)N=1}=0$ and by elimination of
\begin{equation}
\label{equality}
\int_0^1{\rm d}x\,{C}^{(2),(+)}(x,1)=\int_0^1{\rm d}x\,{C}^{(2),(-)}(x,1)
\end{equation}
we obtain
\begin{equation}
\label{moment1}
{C}_{2}^{(\ell)N=1}=2\times 3\int_0^1{\rm d}x\,{C}^{(2),(-)}(x,1)\;.
\end{equation}
Noting that the ${C}^{(2),(-)}(x,1)$ term in
Ref.\cite{vanNeerven:1991nn}
is explicitly proportional to $C_F(C_F-C_A/2)$, we are left with
a single integration over $x$, multiplied by this distinctive
non-planar colour structure. Unlike the contributions
from ${C}^{(2),(+)}(x,1)$, this integral is free of singularities
as $x\rightarrow1$, and hence requires no regularization.
The integrand involves trilogarithms, but elementary integration
by parts reduces it to a regular integral whose integrand involves
nothing more complicated than the product of dilogarithms and logarithms.
Maple then provided a speedy evaluation of the numerical coefficient of
$C_F(C_F-C_A/2)$ to 20 significant figures, for which we readily
found a simple fit with a rational linear combination of the
expected structures $\{1,\zeta(2),\zeta(3),\zeta(4)\}$.
Increasing the accuracy of integration to
30 significant figures we confirmed, with overwhelming confidence,
the analytical form
\begin{eqnarray}
\label{C2}
C_2^{(\ell)N=1}&=&\left[-\frac{141}{32}+\frac{21}{4}\zeta(2)
-\frac{45}{4}\zeta(3)+12\zeta(4)\right]C_F(C_F-C_A/2)\\
\nonumber&\approx&
3.69439249494141137892516966638~C_F(C_F-C_A/2)
\;,
\end{eqnarray}
which validates the first 3 significant figures of the approximate
terms of Eq.~(\ref{result}), obtained in Ref.\cite{Kataev:2003en}
by the far more difficult procedure of evaluating
an integral in Eq.~(\ref{moment}) that has three apparently distinct
colour factors and requires delicate regularization
at the singular endpoint, $x=1$, of the ${C}^{(2),(+)}(x,1)$ function,
interpreted as a distribution.

We now interpret the vanishing of the one-loop corrections to
the anomalous dimension and coefficient function of the $N=1$ non-singlet moment of
charged-lepton--nucleon DIS structure functions as a simple consequence
of the vanishing of all radiative corrections to the Adler sum rule
and the absence of a non-planar one-loop diagram that distinguishes
charged-lepton scattering from neutrino scattering. As already
remarked, this makes the two-loop anomalous dimension coefficient
$\gamma_1^{N=1}$ and the two-loop correction $C_2^{(\ell)N=1}$
scheme-independent. The first place that scheme-dependence may appear
is in the three-loop anomalous dimension coefficient $\gamma_2^{N=1}$,
which appears in Eq.~(\ref{AD}) at order $\alpha_s^2$,
albeit divided by $\beta_0$.
This contribution is in the process of calculation (see for example
Ref.\cite{Moch:2002sn}). We expect its contribution to be small in the
$\overline{{\rm MS}}$-scheme, for reasons discussed in Ref.\cite{Kataev:2003en},
based on experience of next-to-next-to-leading order fits~\cite{Kataev:1997nc}
to the data on $xF_3$ in $\nu N$ DIS from the CCFR collaboration.

Moreover we offer our first conjecture, which is that the 6 possible colour
structures in the three-loop term $\gamma_2^{N=1}$ will occur
only in those 3 combinations suppressed in the large-$N_c$ limit,
namely $C_F^2(C_F-C_A/2)$, $C_FC_A(C_F-C_A/2)$ and $C_F(C_F-C_A/2)N_F$.
If this guess turns out to be wrong, then much of our subsequent discussion
will become questionable. It should be noted that this conjecture applies
exclusively to the $N=1$ moment of the non-singlet charged-lepton structure function
$F_2$. We derive it from the wider hypothesis that the {\em differences}
between non-singlet moments of $F_2$ in charged-lepton scattering and neutrino
scattering will continue to exhibit non-planar suppressions,
beyond the two-loop order at which we have observed them.
Then the suppression of $\gamma_2^{N=1}$ in charged-lepton
scattering at large $N_c$ becomes a special consequence of
the complete vanishing of radiative corrections to the Adler sum rule.

We also note how quickly the two-loop corrections
change their colour structure when one considers moments with $N>1$.
For example the ratio
\begin{equation}
\label{ratio}
R_2^N\equiv{C_2^{(\ell)N}-6C_2^{(\nu)N}\over C_2^{(\ell)N}+6C_2^{(\nu)N}}
={\int_0^1{\rm d}x\,C^{(2),(-)}(x,1)x^{N-1}\over
\int_0^1{\rm d}x\,C^{(2),(+)}(x,1)x^{N-1}}
\end{equation}
is forced to take the value $R_2^{N=1}=1$ at $N=1$, by virtue of the
vanishing of radiative corrections to the Adler sum rule. But
for $N=2$, we obtained from Ref.\cite{Moch:1999eb} the ratio
\begin{equation}
\label{ratio2}
R_2^{N=2}=-{0.505931104\over5.4183241N_c^2 -4N_cN_F - 8.4480127}
\end{equation}
which is negative and small in magnitude at large $N_c$,
and also at $N_c=3$ with $N_F=3$ active flavours, where it takes the value
$R_2^{N=2}=-0.117197668$. Moreover the magnitude of $R_2^N$ continues
to decrease very rapidly with the moment-number, $N$, because the integral in
the numerator of Eq.~(\ref{ratio}) has an integrand that is strongly
suppressed as $x\rightarrow1$. Similarly, we expect the currently known
results for the charged-lepton anomalous dimension $\gamma_2^{N}$,
at several even values of $N$, to give little guidance as to the
eventual value at $N=1$, which must be obtained by analytic continuation of
a complete set of even-$N$ results.

\subsection{Planar approximation, renormalons and $1/Q^2$ corrections}

The limit $N_c\rightarrow \infty$ and the $1/N_c$-expansion
\cite{'tHooft:1973jz} are known to be rather useful for
analyzing the non-perturbative structure of QCD.
Here we will use this framework to characterize our
conjecture about the perturbative corrections
and then seek a non-perturbative consequence.

To do this, we use the planar approximation
formulated in Ref.\cite{Maxwell:1997fg}.
In this approximation one retains, after extracting an overall factor of
$C_F$, only those terms at order $(\alpha_s/\pi)^n$
that contain the leading $N_c$ behaviour for each possible power
of $N_F$. In the case of the order $(\alpha_s/\pi)^n$ contribution
to the coefficient function of Eq.~(\ref{coeff})
this prescription then amounts to selecting
\begin{equation}
\label{planar}
{C}_{n}^{(\ell)N=1}{|}_{\rm{planar}}=
{C}_{F}\sum_{i=0}^{n-1}{\cal{C}}^{(\ell)N=1}_{n,i}
{N}_{F}^{n-1-i}{N}_{c}^{i}\;,
\end{equation}
where the ${\cal{C}}^{(\ell)N=1}_{n,i}$ are pure numbers.
By definition, the planar approximation differs from reality
by (at most) terms of order $1/{N}_{c}^{2}$.
So far we have seen that ${\cal{C}}^{(\ell)N=1}_{1,0}=0$,
since there is no one-loop correction to the coefficient function,
and that ${\cal{C}}^{(\ell)N=1}_{2,1}={\cal{C}}^{(\ell)N=1}_{2,0}=0$,
since only the colour structure $C_F(C_F-C_A/2)=-\frac12C_F N_c^{-1}$
survives at two-loop order in this moment, because of the vanishing of all
radiative corrections to the Adler sum rule and the appearance of a non-planar
factor in the difference between charged-lepton and neutrino structure
functions at two loops. Now let us analyze the consequences
of the rather strong conjecture that the planar approximation~(\ref{planar})
also vanishes at all orders $n>2$.

In general, when it is non-vanishing, a planar approximation
provides us with information in two distinct limits, namely
in the large-${N}_{c}$ limit and also in the
large-$N_F$ limit. The intriguing link that it provides between
these limits is underwritten by the way the large-order behaviour of
perturbation theory is built by renormalon singularities,
as was considered in QCD in the pioneering work
of Ref.\cite{Zakharov:1992bx} and reviewed in detail
in Ref.\cite{Beneke:1998ui}.
This leads one to expect that the asymptotic behaviour
of terms in perturbation theory in $n$th order is of the form
${C}_{n}\sim{K}{\beta}_{0}^{n}{n}^{\delta}{n}!$
(where $\beta_0$ is the first coefficient
of the QCD $\beta$-function)
and so it is natural to develop perturbative coefficients as an expansion in
powers of ${\beta}_{0}$.
The planar approximation is indeed polynomial in ${\beta}_{0}$
and hence can be rewritten as
\begin{equation}
\label{b0nc}
{C}_{n}^{(\ell)N=1}{|}_{\rm{planar}}
={C}_{F}\sum_{i=0}^{n-1}{\tilde{\cal{C}}}_{n,i}^{(\ell)N=1}
{\beta}_{0}^{n-1-i}{N}_{c}^{i}\;,
\end{equation}
where again the ${\tilde{\cal{C}}}^{(\ell)N=1}_{n,i}$ are pure numbers.
This expansion is closely related to the procedure of naive nonabelianization
(NNA) or large-${\beta}_{0}$ approximation
proposed in Refs.\cite{Broadhurst:1994se,Lovett-Turner:1995ti}
in which one replaces $N_F$ by $(11{N}_{c}-3{\beta}_{0})/2$
(for recent applications see
Refs.\cite{Broadhurst:2000yc,Broadhurst:2002bi}). The expansion
of Eq.~(\ref{b0nc})
in ${N}_{c}/{\beta}_{0}$ can be regarded as involving different numbers
of effective
renormalon bubble chains involving powers of
${\beta}_{0}$~\cite{Beneke:1998ui},
inserted in planar diagrams~\cite{Maxwell:1997fg}.
There is a related expansion in ${N}_{F}/{\beta}_{0}$
which is obtained by replacing ${N}_{c}$
by $(3{\beta}_{0}+2N_F)/11$~\cite{Lovett-Turner:1994je,Lovett-Turner:1995ti}
\begin{equation}
{C}^{(\ell)N=1}_{n}{|}_{\rm{planar}}=
{C}_{F}\sum_{i=0}^{n-1}{\hat{\cal{C}}}_{n,i}^{(\ell)N=1}
{\beta}_{0}^{n-1-i}{N}_{F}^{i}\;,
\end{equation}
and here again the ${\hat{\cal{C}}}_{n,i}^{(\ell)N=1}$ are pure numbers.
This expansion, which has been termed the ``dual NNA'', has no direct
Feynman diagrammatic interpretation, but turns out to be rather useful in
making estimates of perturbative corrections to various physical quantities
(see for example Ref.\cite{Broadhurst:2000yc}).

We now consider how the planar approximation is related to renormalon
singularities. Following the work of Parisi~\cite{Parisi:1978bj}
one expects that there will be singularities
in the Borel transforms of QCD observables.
We stress that we are focusing here on a coefficient function, say $C$,
and ignoring any anomalous dimension part, since the
latter will not contain renormalon effects~\cite{Mikhailov:1998xi}.
$C$ will have a
Borel representation
\begin{equation}
C(a)=\int_0^\infty{\rm d}z\,{e}^{-z/a}B[C](z)\;.
\end{equation}
Here $a\equiv{\alpha}_{s}/{\pi}$ and $B[C](z)$ is the Borel transform.
The work of Parisi implies that one
expects branch point singularities in $z$ along the real axis at positions
$z=\pm{z_n}$ where
${z}_{n}\equiv
4n/{\beta}_{0}$,
$n=1,2,3,4,{\ldots}$. Those on the positive real axis are referred to as infrared
renormalons (${\rm IR}_{n}$), and those
on the negative real axis as ultraviolet renormalons (${\rm UV}_{n}$).
Near each of these singularities one expects
the structure
\begin{equation}
B[C](z)=\sum_{i}\frac{{K_i}+O(1{\pm}z/{z}_{n})}
{{(1{\pm}z/{z}_{n})}^{{\delta}_{i}}}\;,
\end{equation}
where the sum is over the contributions of various operators,
and the ${\delta}_{i}$ exponents
depend on their anomalous dimensions.
The large-order asymptotic behaviour of the perturbation theory will
be determined by the dominant
renormalon singularity nearest the origin, and its corresponding operator
with largest ${\delta}_{i}$. The
analysis has been carried out for the Adler $e^{+}e^{-}$-annihilation function,
and for moments of the DIS structure functions
$F_1$, ${F}_{2}$ and ${F}_{3}$, in
Ref.\cite{Beneke:1997qd}. ${\rm UV}_{1}$ gives the
dominant contribution for the Adler $e^{+}e^{-}$-annihilation function,
and contributes, together with ${\rm IR}_{1}$, to the moments of DIS structure
functions.
The same dimension-six operator gives the dominant contribution to
${\rm UV}_{1}$ in all the cases considered.
In the planar approximation one finds the exponent~\cite{Maxwell:1997fg}
\begin{equation}
{\delta}_{+}=2-\frac{{\beta}_{1}}{{\beta}_{0}^{2}}
+\frac{2{N}_{F}}{3{\beta}_{0}}+
\frac{\sqrt{16{N_F}^{2}/9+9{N}_{c}^{2}}}{2{\beta}_{0}}
-\frac{3{N}_{c}}{2{\beta}_{0}}\;,
\end{equation}
and one obtains the asymptotic large-order behaviour for the coefficient
function of the $N$th non-singlet moment of ${F}_{2}$
\begin{equation}
{C}_{n}^{N}\approx{K}_{N}
{\left(\frac{-{\beta}_{0}}{4}\right)}^{n}{n}^{{\delta}_{+}-1}{n}!\;.
\end{equation}
In the large-$N_c$ limit one finds the asymptotic behaviour,
\begin{equation}
{C}_{n}^{N}\approx{K}_{N}
{\left(\frac{-11}{12}\right)}^{n}{N}_{c}^{n}{n}^{19/121}{n}!\;.
\end{equation}
Only the overall constant ${K}_{N}$ depends on the moment taken;
the remaining $n$-dependence is
universal~\cite{Beneke:1997qd}.
Notice that in fact the same $n$-dependence also applies to the moments of
${F}_{1}$ and $F_3$~\cite{Beneke:1997qd}.

Our present conjecture is that the non-singlet moments of $F_2$
in charged-lepton DIS and in neutrino DIS have essentially the {\em same}
planar approximation, as a consequence of some generalization of
the Cutkosky rules that were investigated to two-loop order
in Ref.\cite{vanNeerven:1991nn}.
One obvious consequence is that ${K}_{1}=0$ for the Gottfried sum rule,
since clearly there are no corrections to the Adler sum rule.
For higher moments the ${K}_{N}$ $(N>1)$ will be nonzero, but very simply
related. At $n=2$ loops, one sees from Eqs.~(\ref{moment}) and~(\ref{nu})
that both the ${\ell}$ and ${\nu}$
non-singlet ${F}_{2}$ moments are dominated by ${C}^{(2),(+)}(x,1)$,
at large $N_c$. If it remains true beyond two-loop order that only
the $(+)$ component receives a contribution
from planar diagrams, then one would expect that
$6{{C}_{n}^{(\nu)N}|}_{\rm{planar}}={{C}_{n}^{(\ell)N}|}_{\rm{planar}}$
with the factor of $6$ simply resulting from the normalization of the
Adler and Gottfried sums rules in the most naive quark-parton model.
Not only would we expect
$6C_n^{(\nu)N}-C_n^{(\ell)N}$ to be suppressed by a factor of $1/N_c^2$,
but also to decrease rapidly with the moment number, $N$,
as is the case at two-loop order.

So far we have considered only the leading UV renormalon contribution.
One may anticipate that there is an equally important ${\rm IR}_{1}$
contribution, but to compute the corresponding ${\delta}$
one would need the anomalous dimensions of twist-four operators
contributing to the operator product expansion (OPE)
for the non-singlet moments of $F_2$, which are
not known explicitly. The
expectation would, however, be that the corresponding constant
${K}_{N}^{\rm IR}$
would vanish for $N=1$,
and for $N>1$ should differ by a factor of 6
for the $\nu$ and $\ell$ DIS moments.

Since the leading $1/{Q}^{2}$ OPE corrections
to the moments of DIS structure functions are connected with the leading
${\rm IR}_{1}$ renormalon (for a review, see Ref.\cite{Beneke:1998ui}),
we thus expect higher-twist contributions to the Gottfried sum rule
to be suppressed by a factor
of $\alpha_s/(\pi N_c)\sim1/(N_c^2\log(Q^2/\Lambda^2))$ as $N_c\rightarrow
\infty$,
relative to comparable effects in the Bjorken sum
rules~\cite{Broadhurst:1993ru,Broadhurst:2002bi},
because in the Gottfried sum rule a renormalon chain starts to develop
only in a non-planar three-loop diagram, while in the case
of the Bjorken sum rules it starts to develop in a two-loop planar diagram.

\section{The nucleon sea at large $N_c$}

The previous discussion leads us to believe that
the naive quark-parton model expression for the
Gottfried sum rule, namely $I_G=\frac13$, is not modified by
perturbative effects, or by their resummations as renormalon chains
generating higher-twist effects, in the large-$N_c$ limit.
But in the real world, at $N_c=3$,
the experimental data of the NMC collaboration (see Eq.~(\ref{NMC}))
show a very significant discrepancy from the naive expectation of
$\frac13$.

There are several ways out of this puzzle. One is to say that
$1/N_c^2=1/9$ is not small enough for our considerations to be relevant.
Another is to say that the $1/N_c^2$ suppression to two-loop order
was an accident that will not be repeated at higher loops. To our minds,
the most interesting response is to allow that $1/9$ may be a small
enough factor to take seriously, and that such a
suppression of radiative corrections may persist beyond two loops
and hence be reflected in a suppression of higher-twist corrections,
associated with IR renormalons. Then that
leaves the failure of the naive
Gottfried sum rule to be explained by an intrinsically
non-perturbative flavour asymmetry of the nucleon sea
that is inaccessible to renormalon analysis but
should still be apparent in the $N_c\rightarrow\infty$ limit,
to which we have appealed in our perturbative conjectures and their resummations.

It was interesting to learn from the authors of Ref.\cite{Pobylitsa:1998tk}
that this is indeed the distinctive feature of a chiral-soliton model
based on the work of Ref.\cite{Diakonov:1996sr}.
Briefly, their large-$N_c$ picture, at a very low
normalization point, around $0.6$~GeV, is as follows. Isosinglet
unpolarized distribution functions are large, since they give rise to
sum rules that are proportional to $N_c$; isovector unpolarized distribution
functions appear only at next-to-leading order in $1/N_c$, with
the Adler sum rule satisfied in the form
\begin{equation}
\frac12I_A=1=\int_{-1}^1{\rm d}x\left(u(x)-d(x)\right)
\end{equation}
where the integrand at $x>0$ corresponds to a ``constituent'' quark
contribution and at $x<0$ to an antiquark contribution coming from
$u(x)-d(x)=-\left(\overline{u}(-x)-\overline{d}(-x)\right)$.
The failure of
the Gottfried sum rule at large $N_c$ is attributed to the integral
\begin{equation}
\frac12(3I_G-1)=-\int_{-1}^0{\rm d}x\left(u(x)-d(x)\right)
=\int_0^1{\rm d}x\left(\overline{u}(x)-\overline{d}(x)\right)=O(N_c^0)
\end{equation}
which measures the flavour asymmetry of the nucleon sea at this very low
normalization point. Values of $I_G$ between 0.219 and 0.178 were obtained
for a range of constituent quark masses between 350 and 420 MeV, in fair
agreement with $I_G^{\rm exp}=0.235\pm0.026$ at $Q^2=4\,{\rm GeV}^2$.
Note, however, that the NMC data are at a substantially higher momentum
scale than can be accessed directly by the chiral-soliton model.
For that reason, the authors also compared their predictions for
$\overline{u}(x)-\overline{d}(x)$ with the parton distributions
of Ref.\cite{Gluck:1994uf}, which were initialized at a comparably low scale.
Here too, they claim fair agreement.

There are, of course, several other approaches to the problem
of estimating the light-quark flavour asymmetry of the nucleon sea,
based on meson-cloud models, instanton models and other considerations
(see the reviews of Refs.\cite{Kumano:1997cy,Garvey:2001yq}
and the recent work in Ref.\cite{Karliner:2002pk}).
We have highlighted the results of the chiral-soliton model because it is
based on the large-$N_c$ expansion, used throughout this work.

\section{Conclusions}

Within the large-$N_c$ expansion
we have made the following conjectures, based on rather limited
two-loop input:
\begin{enumerate}
\item Within the framework of light-flavour symmetry,
the radiative corrections to the Gottfried sum rule
are suppressed by a factor $1/N_c^2$, relative to the typical
expectation $O((N_c\alpha_s/\pi)^n)\sim1/(\log(Q^2/\Lambda^2))^n$ at $n$ loops.
We base this on the facts that they vanish at the one-loop level and are merely
of order $(\alpha_s/\pi)^2\sim1/(N_c\log(Q^2/\Lambda^2))^2$ at $n=2$ loops.
\item We expect the unknown three-loop anomalous-dimension
coefficient $\gamma_2^{N=1}$
to be restricted to only 3 of 6 possible colour structures,
namely $C_F^2(C_F-C_A/2)$, $C_FC_A(C_F-C_A/2)$ and $C_F(C_F-C_A/2)N_F$.
\item We expect the ratio of the non-singlet
moments, with $N>1$, for the charged-lepton--nucleon and neutrino--nucleon
$F_2$ structure functions, to maintain the naive ratio 6:1, at large $N_c$,
within the framework of light-quark symmetric perturbative QCD,
after one discounts quark-loop terms involving $N_Fd^{abc}d_{abc}/N_C$, 
which will contribute to the neutrino--nucleon moments.
We have exposed the behaviour $6C_n^{(\nu)N}/C_n^{(\ell)N}=1+O(1/N_c^2)$
for all $N>1$ at $n=2$ loops and expect it to persist at higher loop orders
in the quenched approximation, $N_F\rightarrow0$.
\item Moreover, even at finite $N_c$, we expect this ratio to tend
to unity at high moment-number $N$, as is the case at two loops.
\item We expect higher-twist corrections, of order $1/Q^2$, to follow
the same patterns and hence to be negligible in the Gottfried sum rule
at large $N_c$.
\item In attempting to reconcile this large-$N_c$ perturbative picture
with the significant discrepancy between the measured value for the Gottfried
sum rule and the naive expectation of $\frac13$, we note with interest
the low-energy picture of the nucleon as a chiral soliton
in the large-$N_c$ limit, which leads to an intrinsically
non-perturbative flavour
asymmetry of the nucleon sea~\cite{Pobylitsa:1998tk}.
We believe that current phenomenological
analyses which incorporate a flavour-asymmetric sea as non-perturbative
input, as for example in the most recent parton distributions
of Refs.\cite{Gluck:1998xa,Pumplin:2002vw,Martin:2002aw,Alekhin:2002fv},
capture the essence of this situation, in a manner that
cannot be achieved by radiative corrections,
or by their resummations in the form of higher-twist effects.
\end{enumerate}

{\bf Acknowledgements}
We are most grateful to W.~van Neerven and to C.~Weiss, for clarifying remarks.
DJB and ALK thank the members of IPPP, Durham, UK,
for hospitality during the process of these studies. ALK's work was
done within the program of RFBR Grants N 02-01-00601, 03-02-17047
and 03-02-17177.

{\bf Note added in proof}
Shortly after we submitted our paper, an impressive determination 
of three-loop non-singlet splitting functions appeared in 
Ref.\cite{Moch:2004pa}. 
Using that work, we are now able to determine the three-loop 
anomalous-dimension coefficient 
$\gamma_2^{N=1}\equiv-2\int_0^1{\rm d}x P_{\rm ns}^{(2)+}(x)$, 
with $P_{\rm ns}^{(2)+}(x)$ given by Eq.~(4.9) of Ref.\cite{Moch:2004pa}.
To evaluate it, we note that the corresponding integral of the splitting 
function $P_{\rm ns}^{(2)-}(x)$ of Eq.~(4.10) of Ref.\cite{Moch:2004pa} 
vanishes and hence that 
$\gamma_2^{N=1}=2\int_0^1{\rm d}x[P_{\rm ns}^{(2)-}(x)-P_{\rm ns}^{(2)+}(x)]$
indeed has the colour structure that we anticipated. Performing the integral
analytically, we obtained
\begin{eqnarray*}
\gamma_2^{N=1}&=&
(C_F^2-C_AC_F/2)\bigg\{
C_F\bigg[
290-248\zeta(2)+656\zeta(3)-1488\zeta(4)+832\zeta(5)
\\&&{}
+192\zeta(2)\zeta(3)
\bigg]
+C_A\bigg[
{1081\over9}+{980\over3}\zeta(2)-{12856\over9}\zeta(3)
+{4232\over3}\zeta(4)-448\zeta(5)
\\&&{}
-192\zeta(2)\zeta(3)
\bigg]
+N_F\bigg[
-{304\over9}-{176\over3}\zeta(2)+{1792\over9}\zeta(3)
-{272\over3}\zeta(4)\bigg]\bigg\}\\
&\approx&161.713785 - 2.429260\,N_F
\end{eqnarray*}
by systematic reduction of integrals of harmonic polylogarithms
to Euler sums~\cite{BBBL} with weights up to 5. This
result was checked, to 30 significant figures, by numerical integration
of an integrand involving products of dilogarithms,
obtained after integration by parts. 
Within the framework of light-flavour symmetry,
it leads to radiative corrections
$$
3I_G\approx\left\{\begin{array}{lr}
1+0.035521\,\alpha_s/\pi
  -0.58382\,\alpha_s^2/\pi^2&\mbox{ for }N_F=3\\[6pt]
1+0.038363\,\alpha_s/\pi
  -0.56479\,\alpha_s^2/\pi^2&\mbox{ for }N_F=4
\end{array}\right.
$$
that are even smaller than those estimated in Ref.\cite{Kataev:2003en},
since the anomalous dimension terms of order $\alpha_s^2$ 
cancel about 30\% of the order $\alpha_s^2$ contribution from 
the coefficient function.

\end{document}